\documentclass{article}
\usepackage[colorlinks,linkcolor=red, citecolor=blue]{hyperref}
\usepackage[utf8]{inputenc}
\usepackage{authblk}
\usepackage{setspace}
\usepackage[margin=1.25in]{geometry}
\usepackage{graphicx}
\graphicspath{ {./figures/} }
\usepackage{subcaption}
\usepackage{amsmath}
\usepackage{bm}
\usepackage{hyperref}
\usepackage{url}
\usepackage{xcolor}
\usepackage{lineno}
\usepackage{pifont}
\usepackage[style=nejm, 
citestyle=numeric-comp,
sorting=none]{biblatex}
\title{Atrial Septal Defect Detection in Children Based on Ultrasound Video Using Multiple Instances Learning}

\author[1,2*]{Yiman Liu}
\author[3*]{Qiming Huang}
\author[4*]{Xiaoxiang Han}
\author[5]{Tongtong Liang}
\author[1]{Zhifang Zhang}
\author[1]{Lijun Chen}
\author[3]{Jinfeng Wang}
\author[3]{Angelos Stefanidis}

\author[3$\dag$]{Jionglong Su}
\author[2$\dag$]{Jiangang Chen}
\author[2$\dag$]{Qingli Li}
\author[1$\dag$]{Yuqi Zhang}

\affil[1]{Department of Pediatric Cardiology, Shanghai Children’s Medical Center, School of Medicine, Shanghai Jiaotong University, Shanghai, China}
\affil[2]{Shanghai Key Laboratory of Multidimensional Information Processing, school of communication and electronic engineering, East China Normal University, Shanghai, China}
\affil[3]{School of AI and Advanced Computing, XJTLU Entrepreneur College (Taicang), Xi'an Jiaotong-Liverpool University, Suzhou, China}
\affil[4]{School of Health Sciences and Engineering, University of Shanghai for Science and Technology, Shanghai, China}
\affil[5]{Minhang District Centers for Disease Control and Prevention, Shanghai, China}

\affil[*]{These authors contributed equally to this work.}
\affil[$\dag$]{Authors are equally corresponded.}

\date{}

\begin{document}

\maketitle

%%%%%% Abstract %%%%%%
\begin{abstract}
Purpose: Congenital heart defect (CHD) is the most common birth defect. Thoracic echocardiography (TTE) can provide sufficient cardiac structure information, evaluate hemodynamics and cardiac function, and is an effective method for atrial septal defect (ASD) examination. This paper aims to study a deep learning method based on cardiac ultrasound video to assist in ASD diagnosis. Materials and methods: We select two standard views of the atrial septum (subAS) and low parasternal four-compartment view (LPS4C) as the two views to identify ASD. We enlist data from 300 children patients as part of a double-blind experiment for five-fold cross-validation to verify the performance of our model. In addition, data from 30 children patients (15 positives and 15 negatives) are collected for clinician testing and compared to our model test results (these 30 samples do not participate in model training). We propose an echocardiography video-based atrial septal defect diagnosis system. In our model, we present a block random selection, maximal agreement decision and frame sampling strategy for training and testing respectively, resNet18 and r3D networks are used to extract the frame features and aggregate them to build a rich video-level representation. Results: We validate our model using our private dataset by five-cross validation. For ASD detection, we achieve $89.33\pm3.13$ AUC, $84.95\pm3.88$ accuracy, $85.70\pm4.91$ sensitivity, $81.51\pm8.15$ specificity and $81.99\pm5.30$ F1 score. Conclusion: The proposed model is multiple instances learning-based deep learning model for video atrial septal defect detection which effectively improves ASD detection accuracy when compared to the performances of previous networks and clinical doctors.
\end{abstract}

%%%%%% Main Text %%%%%%

\section{Introduction}

Congenital Heart Defects (CHDs) are the most common birth defects, with an incidence rate of approximately 0.9\% in live births. They are the main causes of death in children aged between 0 and 5 years old \cite{zhao2019prevalence}. The atrial septum separates the left and right atrium during the embryonic period. The abnormal formation of this gap may lead to CHDs after birth. If the gap does not close by itself during growth and development, atrial septal defect (ASD) occurs \cite{chen2022fully}. Based on the occurrence of ASD  can be classified into two categories: primary ASD and secundum ASD. The secundum ASD can be divided into four types: central defect, superior cavity defect, inferior cavity defect, and mixed defect according to the location of the defect \cite{rhodes2002effect}. The types of defect studied in this paper were secundum ASD. ASD in children is one of the most common CHDs, accounting for approximately 6–10\% of CHDs, with an estimated occurrence rate of 1–3 per 1,000 \cite{bradley2020atrial}. The effect of ASD on the heart is a gradual process, starting from the right atrium volume overload to the right atrium expansion, and then to the right ventricle and pulmonary circulation system expansion, thus developing from the atrial shunt to pulmonary hypertension \cite{chen2022fully}. Most children with isolated atrial septal defects are free of symptoms, but the rates of exercise intolerance, arrhythmias, right ventricular failure, and pulmonary hypertension increase with age, and life expectancy is reduced in adults with untreated defects \cite{geva2014atrial}. Therefore, accurate detection of ASD in early childhood is of great clinical importance. \\

Transthoracic echocardiography (TTE) can provide sufficient structural information about the heart and can be used to assess hemodynamics and cardiac function, serving as an effective method for ASD examination. The left and right atrial level shunt signals displayed by color Doppler flow imaging in echocardiography can accurately diagnose ASD \cite{huang2011image}. The advantages of echocardiography include simplicity, non-invasiveness, low cost, repeatability, and accuracy. However, accurate diagnosis by TTE relies heavily on the accurate judgment of echocardiographers, which is time-consuming and subjective \cite{wu2021standard}. In China, especially in the regions with insufficient healthcare, there is a shortage of experienced cardiac ultrasound doctors. Meanwhile, poor-quality and noisy echocardiograms can lead to missed diagnoses and diagnostic errors as echocardiograms are affected by the machine imaging quality. It has been reported that up to 30\% of ultrasound results are not very accurate \cite{lu2009segmentation}. Therefore, the research on the efficient and precise intelligent diagnostic of ASD is of paramount practical significance.

The development of deep learning (DL) \cite{he2016deep}\cite{dosovitskiy2020image}, was soon applied to the field of image analysis \cite{ronneberger2015u}\cite{ronneberger2015u}\cite{yadav2019evalai}\cite{sekuboyina2021verse}. Moreover, there has been much research work on the analysis of ultrasound \cite{lin2022new}\cite{chen2021uscl} and echocardiography \cite{huang2021new}\cite{nagueh2020left}\cite{ostvik2021myocardial}\cite{ahn2023co}. DL was recently shown to achieve exceptional performance in the realm of video understanding. The value of deep learning has been shown in 2D \cite{simonyan2014two}\cite{wang2016temporal}\cite{lin2018nextvlad}\cite{feichtenhofer2019slowfast} and 3D networks \cite{tran2015learning}\cite{carreira2017quo}\cite{tran2018closer}. The model we present in this paper is largely inspired by the aforementioned work in video understanding and offers improvement to it. There are some differences between the video understanding task and the video-based ASD recognition task that this article attempts to solve. Video understanding often involves providing a 3-5 second video and training the network to recognize its semantics. In such cases, the network pays more attention to the time-allowed information between video frames. In the case of video understanding, the model places more focus on the temporal information between video frames. However, during the ASD video recognition task, the video frames of most positive samples only appear for a short period of time. Therefore, how highlighting this segment is an issue that needs to address when designing ASD recognition networks. Furthermore, the review of our experimental results shows that the video classification network based on a 2D model is superior to a 3D model in terms of classification accuracy and efficiency in the case of ASD detection. One plausible reason is that, for echocardiography video, the similarity between frames is quite high, and 3D convolution networks are prone to overfitting this type of data.

 In this study, we offer an atrial septal defect detection algorithm in children based on ultrasound video using multiple instance learning. We propose a block random sampling strategy and maximal agreement decision for training and testing our algorithm. Then, we use resNet18 to extract the frames feature. To enrich the temporal information between different echocardiography frames, we use 3D convolution after stage 2 in resNet18 and generate a temporal representation. Subsequently, for each frame feature, we use attention pooling to aggregate them to form a spatial representation. Finally, we combine the spatial representation and temporal representation to build a general video-level representation for ASD classification. Our novel approach contributes to work in this area in the following ways:

\begin{enumerate}
    \item To the best of our knowledge, classifying ASD based on video-level data is carried out for the first time.
    \item The training and testing video frame sampling techniques we proposed are shown to improve the accuracy of video-based ASD recognition. And these methods are easy to implement and apply to other video classification scenarios.
    \item When  conducting clinical tests that compare our model's capability to classify ASD with that of junior and senior doctors, we conclude that our model outperforms the doctors` results.
\end{enumerate}
 
\section{Materials and methods}

\subsection{Participants}

The subjects of this retrospective study were pediatric patients undergoing color Doppler echocardiography at Shanghai Children's Medical Center from March 1, 2022 to November 1, 2022. These cases included patients diagnosed with ASD (labelled as positive) and normal (labelled as negative).

\subsection{Data Collection} 

This study was approved by the Institutional Review Board of Shanghai Children's Medical Center (Approval No.SCMCIRB-K2022183-1) with a patient exemption applied. All patients were examined via echocardiography using GE Vivid E90 and E95 ultrasound systems with M5Sc and 6S transducers.  We chose two standard views, the subcostal view of the atrial septum (subAS) and the low parasternal four-chamber view (LPS4C) and acquired their color doppler dynamic videos. All dynamic videos were blind reviewed by two senior sonographers and the videos with blurred atrial septum were excluded from the review. Standard imaging techniques were for two-dimensional, M-mode, and color doppler echocardiography in accordance with the recommendations by the American Society of Echocardiography \cite{lopez2010recommendations}. All data was strictly de-identified to protect patient privacy. The original data was stored in the format of Digital Imaging and Communications in Medicine (DICOM) in the Picture Archiving and Communication Systems (PACS) database. For the convenience of program processing, DICOM data is converted into AVI dynamic video.

\begin{figure*}[!h] \centering
  \includegraphics[width=0.8\linewidth]{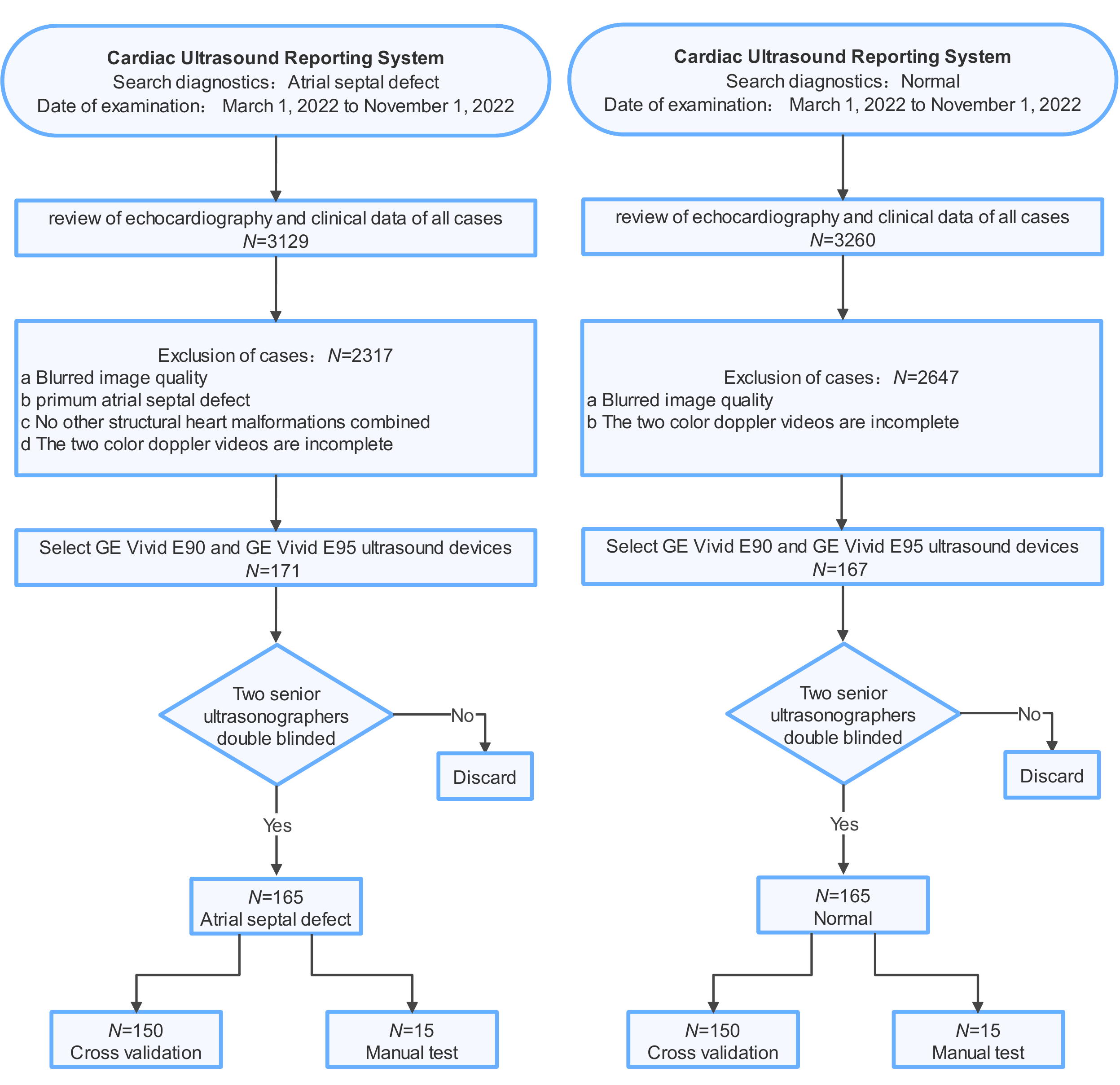} 
  \caption{Inclusion and exclusion criteria for the main cohort of this study}
  \label{inclusion}
\end{figure*}

\begin{figure}[!h] \centering
  \includegraphics[width=0.7\linewidth]{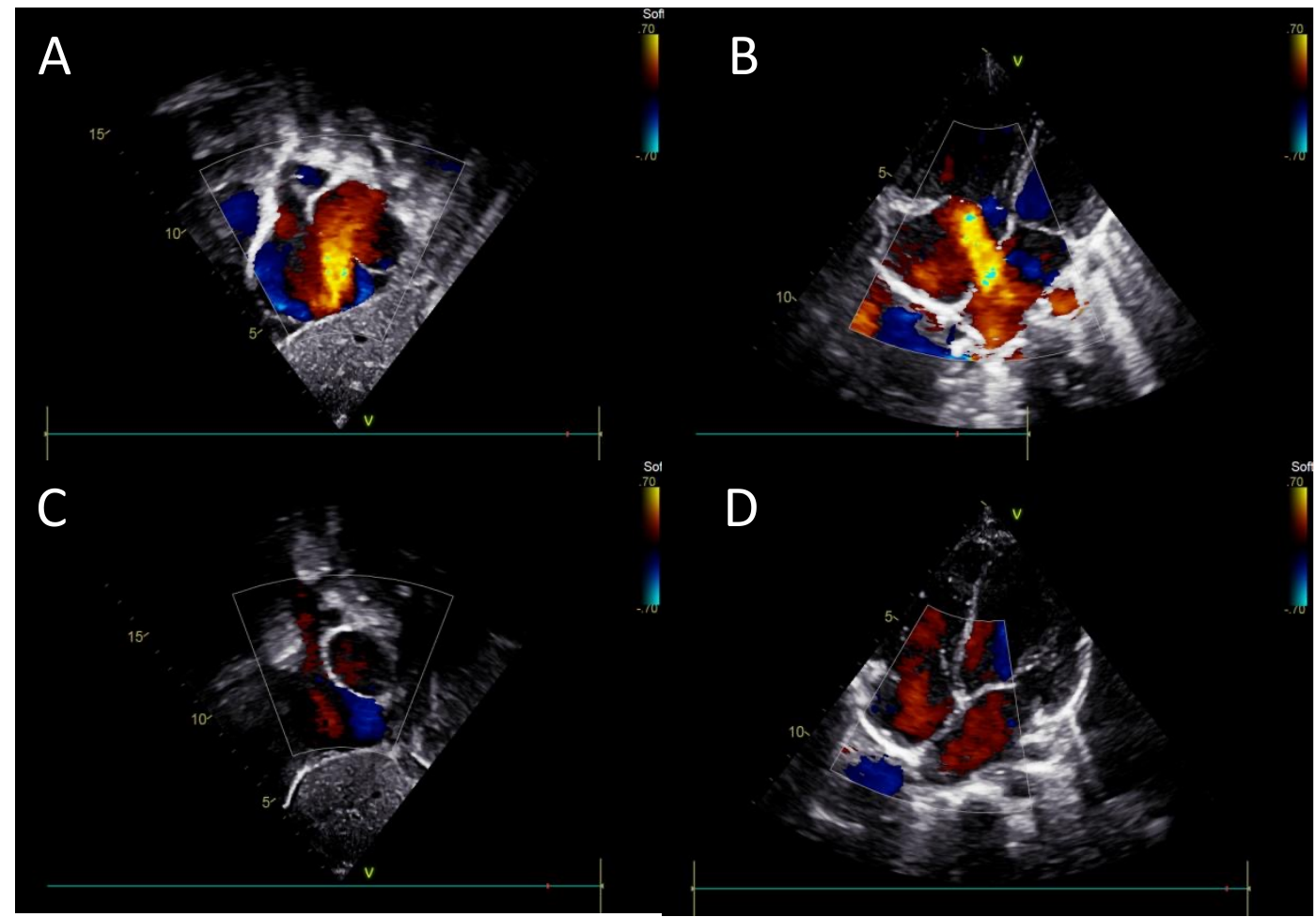} 
  \caption{The visualization of typical ASD and normal static images from the collected dataset. (A) an ASD frame from the subcostal view of the atrial septum, (B) an ASD frame from the low parasternal four-chamber view, (C) a normal frame from the subcostal view of the atrial septum, (D) a normal frame from the low parasternal four-chamber view}
  \label{vis_examples}
\end{figure}

\begin{figure*}[!h] \centering
  \includegraphics[width=\linewidth]{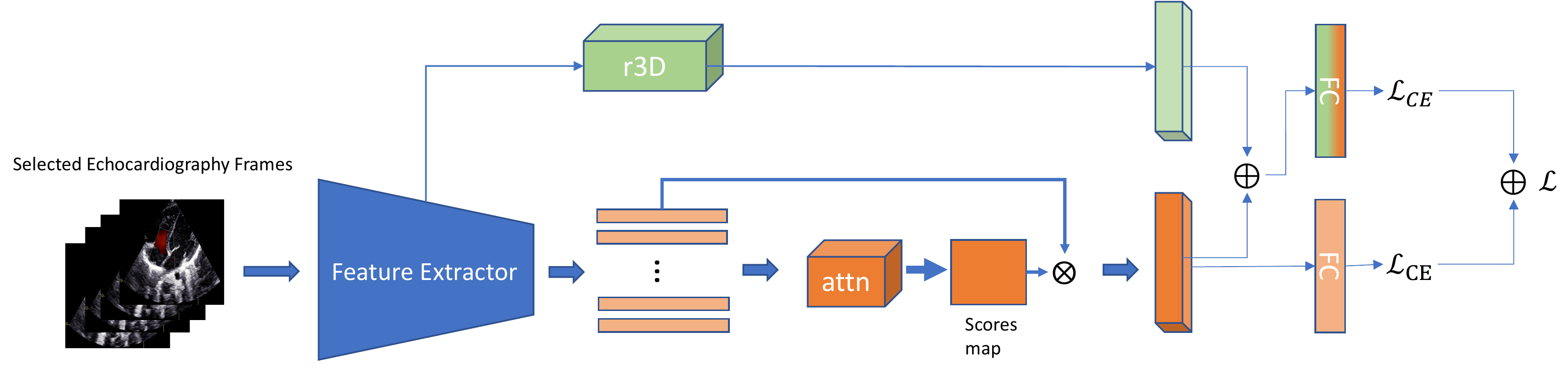} 
  \caption{The architecture of our proposed model for ASD detection. It contains two branches. One branch is employed to extract the spatial features of the video, while the other branch employs an r3D network to capture the temporal features of the video. These two features are then concatenated for final classification.}
  \label{model}
\end{figure*}

\begin{figure*}[!h] \centering
  \includegraphics[width=\linewidth]{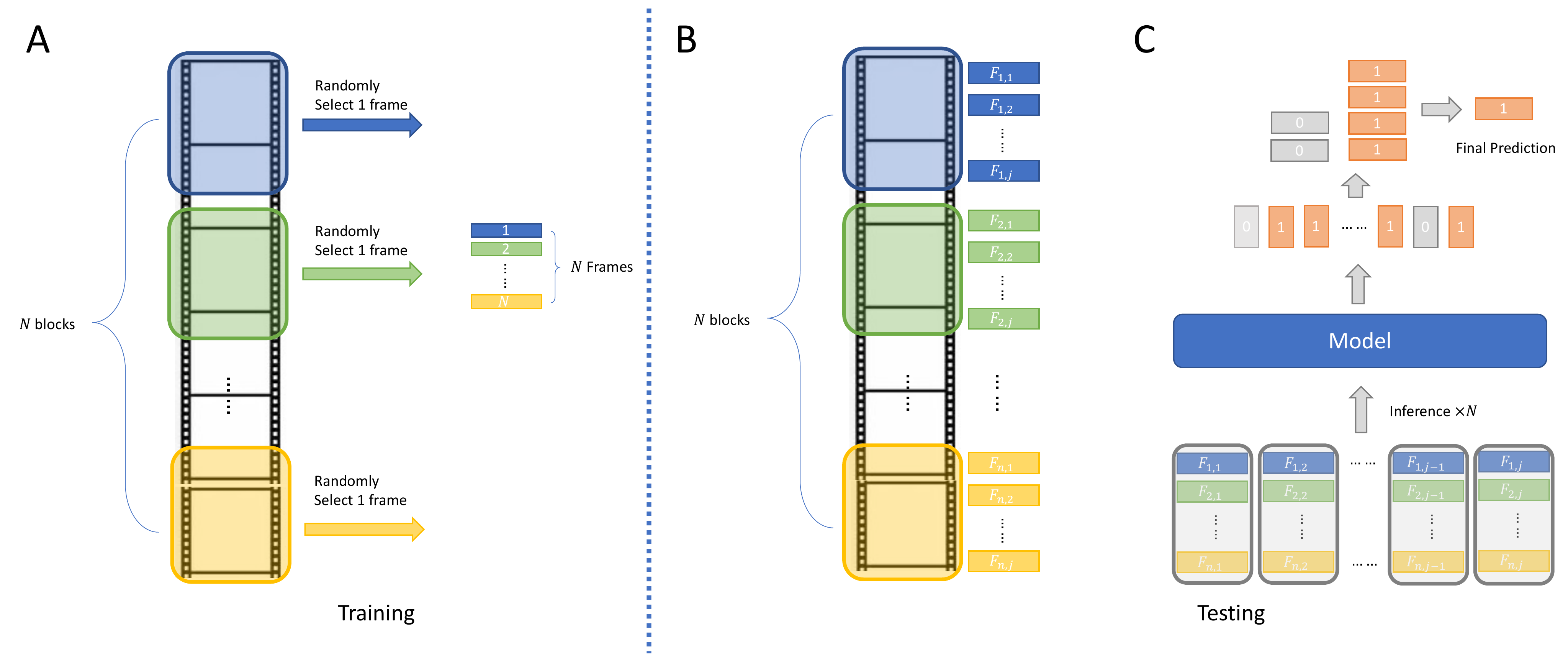} 
  \caption{The proposed frame selection strategies on training and testing stages. (A) Block random selection (BRS) for training. (B) Block inference for testing. We assign each echocardiography frame with a frame index $F_{i,j}$ where $i,j$ indicates the $j^{th}$ frame in $i^{th}$ block. (C) Our proposed maximum agreement decision (MAD) to refine ASD prediction.}
  \label{train_test}
\end{figure*}

\subsection{Training/Validation Dataset}
A total of 300 patients (resulting in 600 dynamic  color doppler videos) are used as the training and validation set for ASD detection, as shown in Fig \ref{inclusion}. The visualization of typical ASD and normal static images from the collected dataset is given in Fig \ref{vis_examples}. The recruitment of patients is initialized from the Ultrasound Report System of The Department of Pediatric Cardiology, Shanghai Children’s Medical Center by retrospective searching. The diagnostic keyword used is set as the  ``Atrial Septal Defect'', and the time of examination is set from March 1, 2022, to November 1, 2022. The obtained cases are reviewed consecutively to evaluate candidacy based on echocardiography results and clinical data. The process is given in Fig. 2. The final number of cases in ASD and normal are 165 and 165 respectively.  While 150 cases in ASD and 150 cases in normal for cross-validation as well as 15 cases in ASD and 15 cases in normal for manual testing.

Specifically, the inclusion criteria for the data collection are: (1) Secundum Atrial Septal Defect; (2) the use of  GE Vivid E90 and GE Vivid E95 ultrasound devices; (3) two color doppler videos of subcostal views of the atrial septum and  low parasternal four-chamber view are available.

A few exclusion criteria are also in place to avoid interference with the training results of the model : (1) blurred image quality; (2) primum atrial septal defect; (3) no other structural heart malformations combined; (4) the two color doppler videos are incomplete.The demographic and clinical characteristics of the training and validation data sets are given in Table \ref{characteristic}.

\begin{table}[!h]\scriptsize \centering
\begin{tabular}{llll}
\hline
\textbf{Characteristics} & \textbf{ASD group ($n$ = 150)} & \textbf{Normal group ($n$ = 150)} & \textbf{$p$-value} \\ \hline
Age (years)              & 2.66(1.41-4.68)            & 2.23(0.91-5.23)                 & 0.15             \\
Female/male              & 88/62                        & 82/68                           & 0.49             \\
Weight (kg)              & 13.50(10.08-17.58)           & 11.10(7.60-18.53)               & 0.06             \\
Height (cm)              & 80.00(93.00-110.00)          & 87.50(69.75-113.25)             & 0.05             \\
Size of ASD (mm)         & 11.1±6.0                    & -                                & -                 \\ \hline
\end{tabular}
\caption{Clinical characteristic comparisons between the ASD group and the normal group of the data set where $n$ denotes the number of cases.}
\label{characteristic}
\end{table}

\subsection{Methods}

In this study, we propose an echocardiography video-based atrial septal defect diagnosis system. The overall model structure is given in Fig \ref{model}. We propose a block random frame sampling strategy for the training stage to train our model robustly. The maximum agreement decision during the testing stage is offered to refine inference prediction. Then, resNet18 is used as a feature extractor to capture the frame features. To enrich the temporal information between different echocardiography frames, we use 3D convolution after stage 2 in resNet18 and generate temporal representation. Afterwards, for each frame feature, attention pooling is used to aggregate them to form a spatial representation. Finally, we combine the spatial representation and temporal representation to build a general video-level representation for ASD classification.

% Our model consists of two branches, a 2D CNN-based spatial feature extractor, and a 3D CNN-based temporal feature extractor. We aggregate these two features to build a rich feature representation for classification. To further improve the robustness of our model as well as its classification accuracy, we propose a block random frame sampling strategy for the training stage to train our model robustly. The maximum agreement decision during the testing stage is offered to refine inference prediction.

\subsubsection{Multiple Instance Learning}

The video-based echocardiography ASD detection problem can be considered as a multiple-instance learning problem. Let $B_i$ denote the $i^{\text{th}}$ bag of frames in echocardiography videos. Let $B_{ij}$ denote the $j^{\text{th}}$ frames in bag $B_{i}, j=1,...,n_{i}$ with a total of $n_i$ frame. Let $\mathcal{F}_{B_{ij}}$ denote the frame feature given by $F(B_{ij})$ where $F(\cdot)$ is deep feature extractor. Let $\mathcal{F}_{B_i}$ denote the video feature given by $\mathcal{G}(\mathcal{F}(B_{ij}))$ where $\mathcal{G}$ is a feature aggregation function. Finally, let $y_i$ denote the label of bag $B_i$, where $y_i\in \{-1, 1\}$. The aim is to establish a linear mapping function with projection $\mathcal{F}_{B_i} \to y_i$.

\subsubsection{Attention Aggregation}

The feature aggregation function is used to represent a bag feature by aggregating all the frame features. Two common methods are $average(\cdot)$ and $max(\cdot)$ where $average(\cdot)$ takes the mean of all frame features to represent a video feature and $max(\cdot)$ takes maximum frame feature to represent a video feature. Because regardless of the feature relationship between different instances, these two methods aggregate features in a fixed manner, lacking flexibility. As a result, we use the attention aggregation module proposed in \cite{ilse2018attention} to aggregate the features. Formally, we let $\mathcal{H}=\{h_1, h_2, ..., h_J\}$ denote a video with $J$ frames. The attention aggregation function is given by:
\begin{equation}
    \mathcal{Z} =\sum_{j=1}^{J}a_j h_j
\end{equation}
where
\begin{equation}
    \alpha_j=\frac{\text{exp}\{\bm{w}^{T}\text{tanh}(\bm{V}\bm{h}_j^{T})\}}{\sum_{j=1}^{J}\text{exp}\{\bm{w}^{T}\text{tanh}(\bm{V}\bm{h}_j^{T})\}}
\end{equation}

% Here \bm{$w$} $\in \mathbb{R}^{L\times 1}$ and \bm{$V$} $\in \ \mathbb{R}^{L\times M}$ are two trainable parameters. 

$tanh(\cdot)$ is a non-linearity activation function. $M$ is the dimension of the mapped feature. We set $M=1024$ instead of  the default value of $512$ in \cite{ilse2018attention} to strengthen the capacity to model complicated data.

\subsubsection{Feature Extraction}

As stated earlier, there are typically two network architectures for video classification: 2D-based and 3D-based. For example, 3D convolutional networks \cite{tran2015learning}\cite{feichtenhofer2020x3d}\cite{tran2018closer} are widely used in video understanding. 3D convolution is used to model the temporal information. However, we argue that in video-based atrial septal defect detection, spatial information is not the principal factor, since a positive frame may appear within a short period of time. Hence, we build our classification model based on the 2D convolutional network. The overall model structure is depicted in Fig \ref{model}. For each frame, we use a resNet18 \cite{he2016deep} to obtain the feature representation and concatenate them to form an $N\times D$ general video representation where $N$ is the number of frames used for training and $D$ is the dimension of the feature. Following this, we use attention pooling to generate $N\times 1$ weights and multiply them by the general feature representation to form the final spatial representation $\mathcal{D}_{spatial}$. Although temporal information is not dominant in ASD detection, we can use 3D convolution \cite{hara2018can} to model the temporal information of color doppler signals. Similar to \cite{zolfaghari2018eco}, we extract the feature map from stage 2 of resNet18 and followed by stages 2 - 4 of r3d to capture temporal information between different training frames and generate a temporal feature representation $\mathcal{D}_{temporal}$. The final representation is given by:
\begin{equation}
    \mathcal{D}= \mathcal{D}_{spatial} + \mathcal{D}_{temporal}
\end{equation}

\subsubsection{Frame Sampling Strategy and Decision Rule}

Since the video frames are computationally more expensive, using all frames in the training stage is time-consuming and inefficient. Furthermore, adjacent video clips have high similarities which may lead to overfitting. To address these, we propose two different frame selection strategies for the training stage and inference stage respectively. Fig \ref{train_test} gives the overall frame sampling method. Similar to \cite{wang2016temporal}, we segment each video into $N$ blocks with each block containing the same number of frames. Subsequently, we randomly select one frame from the whole video to construct an $N$-frames collection used for training. This can be regarded as data augmentation used to improve the robustness of the model. Compared to randomly sampling $N$ frames from the whole video, this method ensures the diversity of frames since we strictly divide the whole video into several blocks and sample frames from each of these blocks. 

Since a positive sample of atrial septal defect may appear in any frames within the video, partially selecting frames may lead to misclassification. Hence, we utilize all frames for inference and propose a decision rule called maximal agreement decision. This decision rule is inspired by the majority rule in human society. Specifically, in the inference stage, we also split the whole video into $N$ blocks, for each block, we select $i$ frame 
, $i \in 1, ..., K$, where $K$ is the number of frames in a video block. $B_i, i\in 1,..., N$ is denoted as the $i$ th frames collection used for testing. We then have $N$ predictions $[Y_1, ..., Y_N], Y_i \in \{-1, +1\}, \forall i \in \{1,2,...,n\}$. The final prediction $\hat{Y}$ for this video is determined by :
\begin{equation}
    \hat{Y} = \max \left\{\sum_{i=1}^{N} Y_i = -1, \sum_{i=1}^{N} Y_i = 1\right\}
\end{equation}

This decision rule mimics the difficult and severe diseases in real clinical practice. We consider $Y_i$ as conclusions from all experts. Then, the major rule is used to obtain the final diagnosis.

% \textcolor{red}{It is intuitive since the minority obeying the majority is a common decision rule}.

% \begin{figure}[!h] \centering
%   \includegraphics[width=0.9\linewidth]{src/mad_.pdf} 
%   \caption{Maximal Agreement Decision Method}
%   \label{mad}
% \end{figure}

% \subsubsection{Loss function}

% We have two loss functions. One is for spatial representation and the other is for temporal representation as an auxiliary loss. The final loss function is calculated by:
% \begin{equation}
%     \mathcal{L} = \mathcal{L}_{spatial} + \mathcal{L}_{temporal}
% \end{equation}

% \begin{figure}[!h] \centering
%   \includegraphics[width=\linewidth]{src/model.pdf} 
%   \caption{Overall structure of proposed method}
%   \label{model}
% \end{figure}

\section{Experiments}

\subsection{Experiments setup}

We implement our model in PyTorch using an NVIDIA TESLA T4 GPU accelerator. The SGD optimizer is used with
an initial learning rate of 1e-4. During training, we resize the video frames to 224$\times$224, we select $N=16$ frames from videos for training with batch size 32. In the main experiment, five-fold cross-validation is used to evaluate the performance of the proposed model. There is no data augmentation used in training.

Three hundred and thirty cases are used in this study. Specifically, three hundred cases are used to assess the performance of the proposed method. The remaining 30 cases are excluded from the training and are directly used for additional testing and their demographic and clinical characteristics of the dataset are shown in Table \ref{characteristic_2}. 

\begin{table}[!h]\scriptsize \centering
\begin{tabular}{llll}
\hline
\textbf{Characteristics} & \textbf{ASD group ($n$ = 15)} & \textbf{Normal group ($n$ = 15)} & \textbf{$p$-value} \\ \hline
Age (years)              & 3.33(1.33-5.74)            & 3.57(1.25-5.58)                 & 0.96             \\
Female/male              & 9/6                        & 8/7                           & 0.72             \\
Weight (kg)              & 16.00(11.50-17.50)           & 16.50(10.50-18.00)               & 0.74             \\
Height (cm)              & 85.00(105.00-110.00)         & 86.00(106.00-112.00)             & 0.86             \\
Size of ASD (mm)         & 13.1±7.4                    & -                                & -                 \\ \hline
\end{tabular}
\caption{Clinical characteristic comparisons between the ASD group and the normal group of the test data set where $n$ denotes the number of cases.}
\label{characteristic_2}
\end{table}

 \subsection{Evaluation protocols}

 We mainly evaluate our proposed model using confusion matrix analysis, described by the following metrics:
\begin{equation}
    \text{Accuracy} = \frac{TP + TN}{TP + TN + FP + FN}
\end{equation}

\begin{equation}
    \text{Sensitivity} = \frac{TP}{TP + FN}
\end{equation}

\begin{equation}
    \text{Specificity} = \frac{TN}{TN + FP}
\end{equation}

\begin{equation}
    F1 = \frac{TP}{TP + \frac{1}{2}(FP + FN)}
\end{equation}

\noindent
where TP is true positive, TN is true negative, FP is false positive and FN is false negative.

\begin{table*}[!h] \centering \small
\begin{tabular}{lllllll}
\hline
Model        & AUC(\%) & Accuracy(\%) & Sensitivity (\%) & Specificity(\%) & F1(\%) \\ \hline

\textbf{X3D\cite{feichtenhofer2020x3d}}              &$70.12\pm1.89$         & $68.02\pm2.33$             & $71.69\pm3.39$                 & $67.85\pm5.12$                            & $69.72\pm2.22$       \\
\textbf{R2plus1D\cite{tran2018closer}}              &$75.21\pm1.03$         & $73.09\pm0.88$             & $77.42\pm1.97$                 & $72.34\pm2.66$          & $78.05\pm1.30$       \\ 
\textbf{ResNet18 \cite{he2016deep}} & $86.35\pm4.00$      &$77.25\pm3.82$        & $77.19\pm2.29$                        & $76.92\pm3.12$      & $77.87\pm0.91$  \\
\textbf{Ours} & \bm{$89.33\pm3.13$}    & \bm{$84.95\pm3.88$}        & \bm{$85.70\pm4.91$} & \bm{$81.51\pm8.15$} & \bm{$81.99\pm5.30$}   \\\hline
\end{tabular}
\caption{The ASD classification result of different models via five five-fold cross-validations (Mean±Std)}
\label{result}
\end{table*}

\section{Results and Discussion}
\subsection{Main Results}
As shown in Table \ref{result}, compared to different models including 3D CNN based, our proposed model achieves the best performance with $89.33\pm3.13$ ACU and $84.95\pm3.88$ accuracy on 5-fold cross-validation. The details of sensitivity, specificity and F1 score are given in Table \ref{result}. In addition, we compare the ASD classification accuracy with doctors` diagnoses of ASD. In 30 cases of testing both junior and senior doctors can achieve ASD detection accuracy of 63.66\% and 71.70\% respectively, while our model achieves 83.33\% accuracy as shown in Table \ref{doctor}.

\subsection{Ablation study}

We also report the ablation study of our components, i.e, 3D fusion, attention aggregation, maximum agreement decision and block random selection. The results, given in Table \ref{ablation_fusion} and Table \ref{ablation_mad}, demonstrate the effectiveness of our proposed model for improving ASD detection accuracy.

\begin{table}[!h]\centering
\begin{tabular}{lllll}
\cline{1-4}
Model    & 3D Fusion         & AAM          & Accuracy(\%)          &  \\ \cline{1-4}
ResNet18 & \ding{55}            & \ding{55}       & $77.25 \pm 3.82$      &  \\
         & \ding{55}            & \ding{52}    & $82.59 \pm 2.83$      &  \\
         & \ding{52}         & \ding{55}            & $80.21 \pm 2.11$ &  \\
         & \ding{52}         & \ding{52}    & $84.06 \pm 3.38$       &  \\\cline{1-4}
\end{tabular}
\caption{Ablation Study for 3D Fusion, attention aggregation module (AAM) via five five-fold cross-validations (Mean$\pm$Std).}
\label{ablation_fusion}
\end{table}

\begin{table}[!h]\small \centering
\begin{tabular}{lllll}
\cline{1-4}
Model & MAD          & BRS          & Accuracy(\%)      &  \\ \cline{1-4}
Our model  & \ding{55}       & \ding{55}       & $84.06\pm3.38$    &  \\
      & \ding{55}       & \ding{52}    & $84.59 \pm 2.83$   &  \\
      & \ding{52}    & \ding{55}       & $84.44 \pm 4.41$   &  \\
      & \ding{52}    & \ding{52}    & $84.95 \pm 3.88$   &  \\ \cline{1-4}
\end{tabular}
\caption{Ablation Study for MAD, BRS via five five-fold cross-validations (Mean$\pm$Std) and our model refers to Resnet18+3D Fusion+AAM.}
\label{ablation_mad}
\end{table}

% \begin{table}[!h]\small \centering
% \begin{tabular}{lllllll}
% \hline
% Model           & 3D Fusion & AAM & MAD & BRS & Accuracy(\%)\\ \hline
% % ResNet-50        &          &    &    & &  $83.33\pm2.68$          \\
% % X3D\cite{feichtenhofer2020x3d}        &          &    &    & &        $68.02\pm2.33$    \\
% % R2plus1D        &          &    &    & &   xxx\pm xxx         \\ \hline
% ResNet18        &   \bm{-}      &  \bm{-}  &  \bm{-}  & \bm{-} & $77.25\pm3.82$           \\
%                 &    \bm{-}      &  \bm{-}  & \ding{52}   & \bm{-} & $79.26\pm3.42$         \\
%                 &   \bm{-}       & \ding{52}   &  \bm{-}  & \bm{-} & $82.59\pm2.83$         \\
%                 % &          & \ding{52}   & \ding{52}   & & ?         \\
%                 & \ding{52}         & \ding{52}   &   \bm{-}  & \bm{-} & $84.06\pm3.38$         \\
%                 & \ding{52}         & \ding{52}   & \ding{52}   & \bm{-} & $84.44\pm4.41$    \\   & \ding{52}         & \ding{52}   & \ding{52}   & \ding{52} & \bm{$84.95\pm3.88$ }  \\ \hline
% \end{tabular}
% \caption{Ablation Study for 3D Fusion, attention aggregation module (AAM), maximum agreement decision (MAD) and block random selection (BRS) via five five-fold cross-validations (Mean$\pm$Std)}
% \label{ablation}
% \end{table}

\begin{table}[!h]\centering
\begin{tabular}{llll}
\hline
Model               & Accuracy(\%) & PPV(\%) & NPV(\%) \\ \hline
Junior Doctor       & 63.66             & 53.49        & 89.47        \\
Senior Doctor & 71.70             & 60.53         & $\bm{91.67}$       \\ 
\textbf{Ours}       & $\bm{83.33}$        & $\bm{85.71}$   & 81.25   \\\hline
\end{tabular}
\caption{ASD detection performance of our model and professional doctors based on additional testing.}
\label{doctor}
\end{table}

\subsection{Discussion}

In Table \ref{doctor}, it can be seen that the accuracy for ASD detection of our proposed model is higher than that of both junior doctors and senior doctors detection rates. However, the NPV of our model is lower than that of the doctors. A similar result regarding early Alzheimer’s disease prediction using deep learning is reported in \cite{hu2023vgg}. Positive Predictive Value (PPV) is an indicator to measure the ability of the model to avoid misdiagnosis and Negative Predictive Value (NPV) is used to measure the missed diagnosis ability. Higher PPV can indicate the potential of our model to use as an ASD assistance tool in clinical.

\begin{figure}[!h] \centering
  \includegraphics[width=0.7\linewidth]{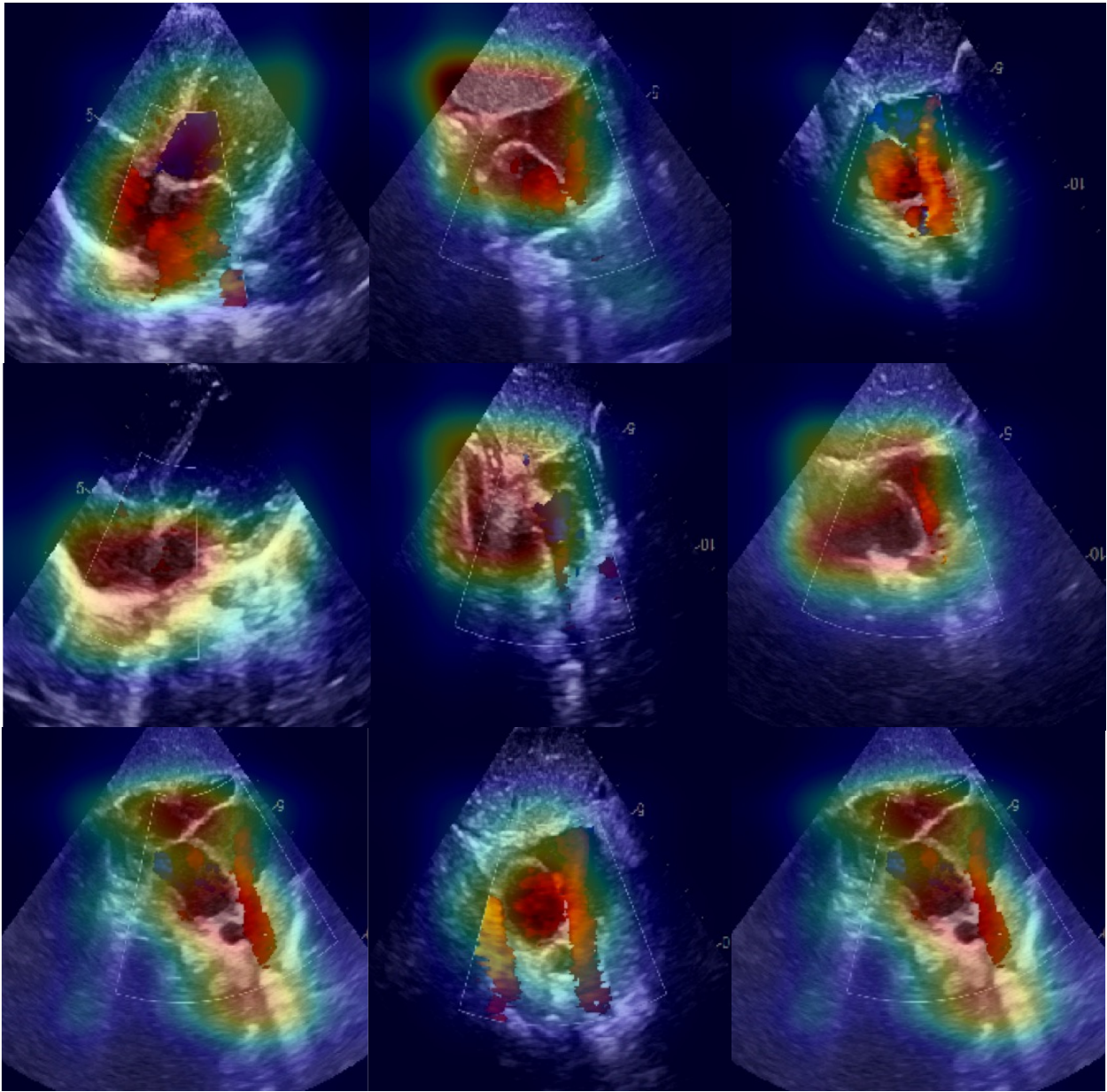} 
  \caption{Visualization of heat map for our model when prediction.}
  \label{vis}
\end{figure}

The heat map of our model is given in Fig \ref{vis}. The most activated areas are shown by the color doppler signal area which indicates the blood flow signal. The importance of the color doppler signal in the diagnosis of ASD is emphasized by this finding.

In addition, we find that ASD shunt blood flow is not present in every frame of the cardiac cycle due to the systole and diastole of the heartbeat. To the best of our knowledge, this study is the first attempt to diagnose ASD in children at the ultrasound video level. It has been reported that deep learning algorithms can guide inexperienced diagnosticians to obtain echocardiograms for limited diagnosis \cite{narang2021utility}. Therefore, our design of ultrasound video diagnosis algorithm of ASD in children is of great clinical importance: for experienced doctors, it can help improve the efficiency of film review and ensure the reliability of ASD diagnosis; for less experienced doctors or even those who do not have the ability to review films, it can help provide echocardiography-based ASD diagnostic information and improve aiding the comprehensiveness of the diagnosis.

% \begin{enumerate}
%     \item Achieving competitive performance compared with \cite{wenjing2022automatic}. 
%     \item Low interpretability.
%     \item Only cases with a defect size of more than 5 mm are selected as their research object\cite{wenjing2022automatic} while some size cases were included in our study.
% \end{enumerate}

\section{Conclusion}

This paper proposes a multi-instance learning-based model for the diagnosis of atrial septal defects using ultrasound video that has achieved excellent results. Since atrial septal defect is a very common cardiac disease that is found in a large number of children, our method can not only help cardiac sonographers to improve the efficiency of diagnosis but also assist primary cardiac sonographers in making diagnostic decisions.

However, the limitations of our study are four-fold: First, we select ultrasound videos from using only two standard views for diagnosing atrial septal defects, while there are some other ultrasound videos from non-standard views that can diagnose atrial septal defects that are not included in our study; Second, our data sets were all secondary foramen ovale atrial septal defects and exclude primary foramen ovale atrial septal defects.  In addition, the shunt direction of the defects is all left-to-right shunts from the left atrium to the right atrium; Finally, the amount of data in our article was small and single-centred. These aspects will be examined in our future work.

\section{Acknowledgement}

% CONFLICT OF INTEREST

% The authors declare that the research was conducted in the absence of any commercial or financial relationships that could be construed as a potential conflict of interest.

% ETHICAL STATEMENT

% Study approval was granted by the Institutional Review Board of Shanghai Children's Medical Center (Approval No:SCMCIRB-K2022183-1). The procedures were performed in accordance with the Declaration of Helsinki and International Ethical Guidelines for Biomedical Research Involving Human Subjects. 

% AUTHOR CONTRIBUTIONS

% Conception and design: YM Liu, QM Huang, XX Han; (II) Administrative support: ZZ Yang, WJ Bao; (III) Provision of study materials or patients: YM Liu JX Zhu; (IV) Collection and assembly of data: YM Liu, QM Huang, XX Han; (V) Data analysis and interpretation: YM Liu, QM Huang, XX Han; (VI) Manuscript writing: All authors; (VII) Final approval of manuscript: All authors.

% FUNDING

This work is supported by the National Natural Science Foundation of China (Grant No. 61975056), the Shanghai Natural Science Foundation (Grant No. 19ZR1416000), the Science and Technology Commission of Shanghai Municipality (Grant No. 20440713100) and the Key Program State Fund in XJTLU (KSF-A-22).

\printbibliography

\end{document}